\documentclass[english]{elsart}
\usepackage[T1]{fontenc}
\usepackage[latin1]{inputenc}
\usepackage{amsmath}
\usepackage{setspace}
\usepackage{amssymb}

\makeatletter

\providecommand{\LyX}{L\kern-.1667em\lower.25em\hbox{Y}\kern-.125emX\@}

\usepackage{setspace}

\usepackage[numbers]{natbib}

\usepackage[dvips]{graphicx}
\usepackage{epic,epsfig}
\usepackage{subfigure,float}
\usepackage{amsmath,amsfonts,amssymb}
\usepackage{inputenc}
\usepackage{natbib}
\usepackage{euscript}

\makeatother
\begin{document}
\begin{frontmatter}

\title{Dynamic asset trees and Black Monday}
\author[label1]{J.-P. Onnela,}
\author[label1]{A. Chakraborti,}
\author[label1]{K. Kaski}
\author[label1,label2]{and J. Kertész}
\address[label1]{Laboratory of Computational Engineering, Helsinki University of Technology, 
P.O. Box 9203, FIN-02015 HUT, Finland}

\address[label2]{Department of Theoretical
Physics, Budapest University of Technology and Economics,
Budafoki út 8, H-1111, Budapest, Hungary }

\begin{abstract}
The minimum spanning tree, based on the concept of ultrametricity, is constructed from the correlation matrix of stock returns. The dynamics of this asset tree can be characterised by its normalised length and the mean occupation layer, as measured from an appropriately chosen centre called the `central node'. We show how the tree length shrinks during a stock market crisis, Black Monday in this case, and how a strong reconfiguration takes place, resulting in topological shrinking of the tree.
\end{abstract}

\end{frontmatter}

The study of the clustering of companies using the correlation matrix of asset returns with a simple transformation of the correlations into distances, producing a connected graph, was suggested by Mantegna \cite{Man}, and later studied by Bonanno et al. \cite{Gio2}. In the graph the nodes correspond to the companies, and the distances between them are obtained from the correlation coefficients. Clusters of companies are identified by means of minimum spanning tree. Many studies have been carried out on clustering in the financial market such as \cite{Kull,Gio,Kas,Lal,Mar}, and on financial market crashes \cite{Lillo,Lillo2}. 
Recently we have studied a set of asset trees obtained by sectioning the return time series appropriately, and determining the minimum spanning trees from correlations between stock returns \cite{jpo}, according to Mantegna's methodology. This multitude of trees was interpreted as a sequence of evolutionary steps of a single `dynamic asset tree'. In addition, we have introduced different measures to characterise the system, such as normalised tree length and mean occupation layer, and they were found to reflect upon the state of the market. The minimum spanning tree, as a strongly pruned representative of asset correlations, was found to be robust and descriptive of stock market events. In this paper, we give a brief demonstration of how the prominent 1987 stock market crash, which culminated in Black Monday (October 19, 1987), may be viewed from the perspective of dynamic asset trees.

First we give a brief description of the methodology. Assume that there are $N$ assets with price $P_{i}(t)$ for asset $i$ at time $t$. The logarithmic return of stock $i$ is given by $r_{i}(t)=\ln P_{i}(t)-\ln P_{i}(t-1)$, which for a certain sequence of trading days forms a return vector $\boldsymbol r_{i}$. In order to characterise the synchronous time evolution of stocks, we use the concept of equal time correlation coefficient between stocks $i$ and $j$, defined as 

\begin{equation}
\rho _{ij}=\frac{\langle \boldsymbol r_{i}\boldsymbol r_{j}\rangle -\langle \boldsymbol r_{i}\rangle \langle \boldsymbol r_{j}\rangle }{\sqrt{[\langle \boldsymbol r_{i}^{2}\rangle -\langle \boldsymbol r_{i}\rangle ^{2}][\langle \boldsymbol r_{j}^{2}\rangle -\langle \boldsymbol r_{j}\rangle ^{2}]}},\end{equation}

where $\left\langle ...\right\rangle $ indicates a time average over the trading days included in the return vectors. The $N\times N$ correlation matrix is transformed to an $N\times N$ distance matrix with elements $d_{ij}=\sqrt{2(1-\rho _{ij})}$, such that $2\geq d_{ij}\geq 0$, respectively. The $d_{ij}$s fulfil the requirements of distances, even ultrametricity \cite{Man}. The distance matrix is then used to determine the minimum spanning tree (MST) of the distances, denoted by $\mathbf{T}$, which is a simply connected graph that links the $N$ nodes with the $N-1$ edges such that the sum of all edge weights, $\sum _{(i,j)\in \mathbf{T}}d_{ij}$, is minimum. It should be noted, that in constructing the minimum spanning tree, we are effectively reducing the information space from $N(N-1)/2$ separate correlation coefficients to $N-1$ separate tree edges.

The dataset we have used in this study consists of daily closure prices for 116 stocks of the S\&P 500 index \cite{sp}, obtained from Yahoo \cite{yahoo}. The time period of this data extends from the beginning of 1982 to the end of 2000, including a total of 4787 price quotes per stock. We divided this data into $M$ \emph{windows} {$t=1,\, 2,...,\, M$ of width $T$ corresponding to the number of daily returns included in the window, where consecutive windows were displaced by $\delta T$. In our study, $T$ was typically set between 2 to 6 years (500 to 1500 trading days) and $\delta T$ to one month (about 21 trading days). 

In order to study the temporal state of the market we defined the \emph{normalised tree length} as $L(t)=\frac{1}{N-1}\sum _{d_{ij}\in \mathbf{T}^{t}}d_{ij}$, where $t$ denotes the time at which the tree is constructed, and $N-1$ is the number of edges in the MST. Figure 1 shows the normalised tree length $L$ as a function of time and window width. The two sides of the ridge converge as a result of extrapolating the window width $T\rightarrow 0$ \cite{jpo}, pointing to Black Monday. In fact, we find that the normalised tree length $L(t)$ decreases during a market crash, indicating that the nodes on the tree are strongly pulled together, i.e. the tree shrinks in length.

In addition to the length, we are also interested in the topology of the tree. The robustness of asset tree topology can be investigated by the \emph{single-step survival ratio} defined as $\sigma _{t}=\frac{1}{N-1}|E^{t}\cap E^{t-1}|$. In this $E^{t}$ refers to the set of edges of the graph at time $t$, $\cap $ is the intersection operator and $|...|$ gives the number of elements in the set. In other words, the survival ratio is the fraction of edges found common in both graphs, which are one time step $\delta T$ apart. Under normal circumstances, the graphs in two consecutive time windows $t$ and $t+1$ should look very similar, at least for small values of $\delta T$. In Figure 2, we have depicted $\sigma _{t}$ for two window width values, where we find two prominent dips indicating that two strong tree reconfigurations take place. These two dips are positioned symmetrically around Black Monday and are, in fact, found to converge into it upon extrapolating $T \to 0$. Thus we have shown that a remarkable change in tree topology takes place at the time of the market crash. In order to characterise this change, a new measure is needed. 

To establish a reference in the graph, we introduced the concept of a \emph{central node}. The central node is central, i.e. important, in the sense that any change in its price strongly affects the course of events in the market as a whole. Two alternative definitions emerged for the central node as either (i) the node with the highest \emph{vertex degree} (number of incident edges), or (ii) the node with the highest correlation coefficient weighted vertex degree. In addition, one can have either (a) static (fixed at all times) or (b) dynamic (continuously updated) central node, without considerable effect on the results. In our studies, General Electric (GE) was chosen as the static central node, since for about 70\% of the time windows it turned out to be the most connected node. In practice, both definitions yield very similar results, independnet of whether static or dynamic central node is employed. In general, roughly 80 percent of the time, the central node coincides with the center of mass of the tree, the exact figure depending on the values of the parameters \cite{jpo1}. 

In addition, we have characterised the tree topology or, more specifically, the location of the nodes in the tree. This is done by defining the \emph{mean occupation layer} as $l(t)=\frac{1}{N}\sum _{i=1}^{N}\mathop {\mathrm{lev}}(v_{i}^{t})$, where $\mathop {\mathrm{lev}}(v_{i})$ denotes the level of vertex $v_{i}$ in relation to the central node, whose level is taken to be zero. In general, the mean occupation layer is found to fluctuate as a function of time (for a plot see \cite{jpo}). However, one can easily identify two sharp dips located symmetrically around Black Monday. Upon extrapolating the window width $T \to 0$, the dips are found to converge and, thus, the tree shrinks in topology during the crash. As one cannot evaluate the graph in the limit, a practical demonstration of the effect for $T=1000$ is provided in Figure 3, where plots of normal and crash market topology are presented. The first one was chosen from a normal, business as usual period, resulting in $l(t_{normal}) \approx 3.1$. The latter corresponds to one of the two peaks of the mean occupation layer around the crash, yielding $l(t_{crash}) \approx 2.1$, thus a value clearly below normal. \cite{jpo}

In summary, we have studied dynamic asset trees with reference to the 1987 stock market crash and, in particular, Black Monday. We have shown that the normalised tree length decreases during the crash. Using the concept of single-step survival ratio, we have found the tree topology to undergo a strong reconfiguration during the crash. We have used the mean occupation layer to characterise the nature of this reconfiguration, and found its value to fall at the time of the market crisis. Thus, the shrinking of asset trees on Black Monday is a twofold phenomenon - the asset tree shrinks in terms of tree length, as well as topology.

This research was partially supported by the Academy of Finland, Research Centre for Computational Science and Engineering, project no. 44897 (Finnish Centre of Excellence Programme 2000-2005) and OTKA (T029985).


\clearpage

\begin{figure}
\begin{center}
\includegraphics[width=1\textwidth]{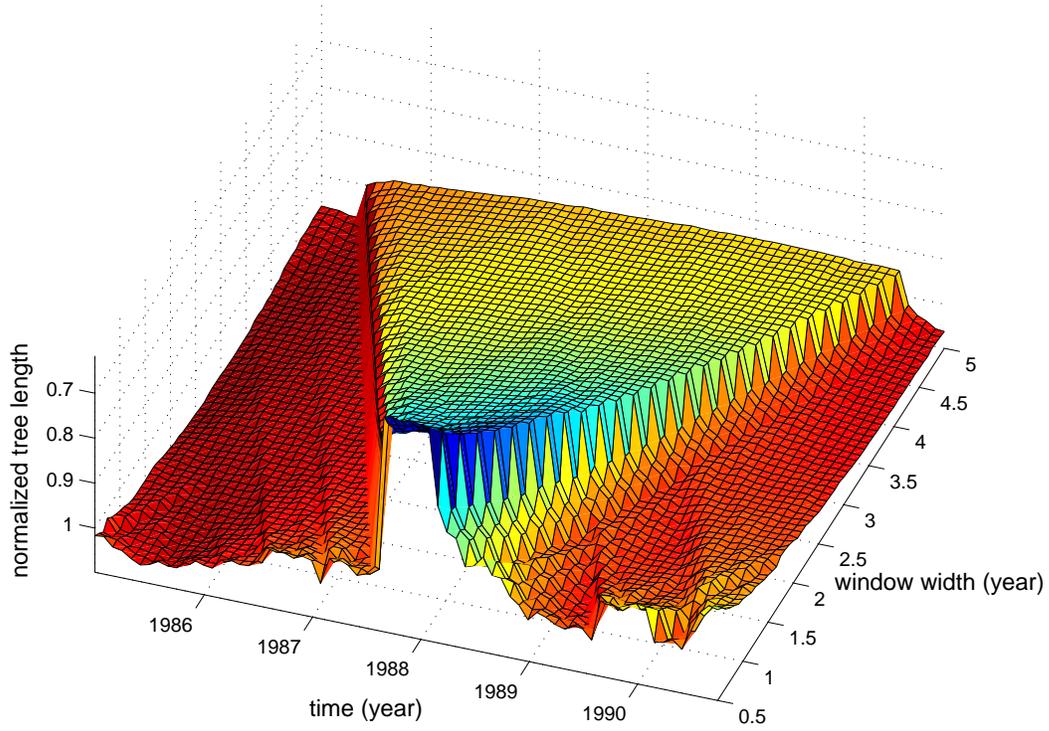}
\end{center}
\caption{Convergence of normalised tree length $L(t)$ as a function of window width. Upon extrapolation the two sides of the ridge are found to converge to Black Monday.}
\end{figure}

\begin{figure}
\begin{center}
\includegraphics[width=1\textwidth]{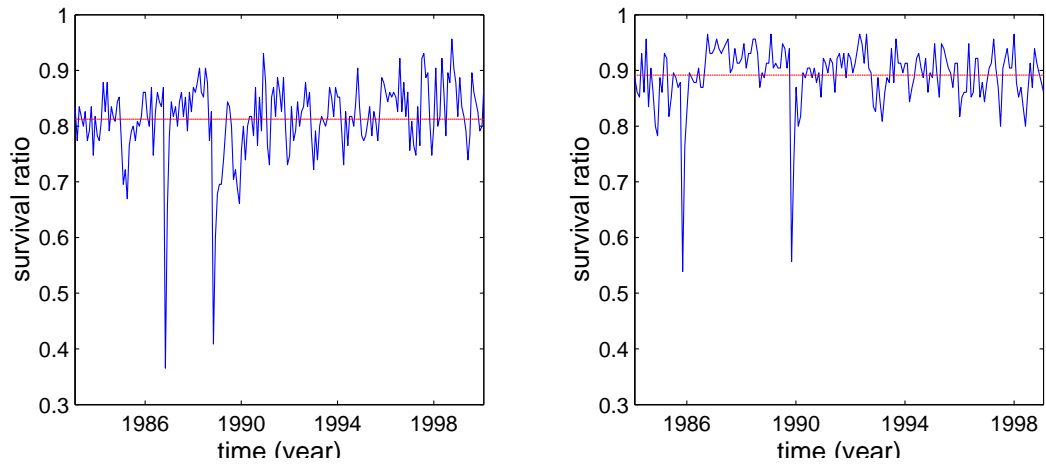}
\end{center}
\caption{Single-step survival ratio $\sigma_t$ as a function of time for $T$=2yrs (left) and $T$=4yrs (right). The prominent dips in both plots indicate that a strong tree reconfiguration takes place.}
\end{figure}

\clearpage

\begin{figure}
\begin{center}
\includegraphics[width=1\textwidth]{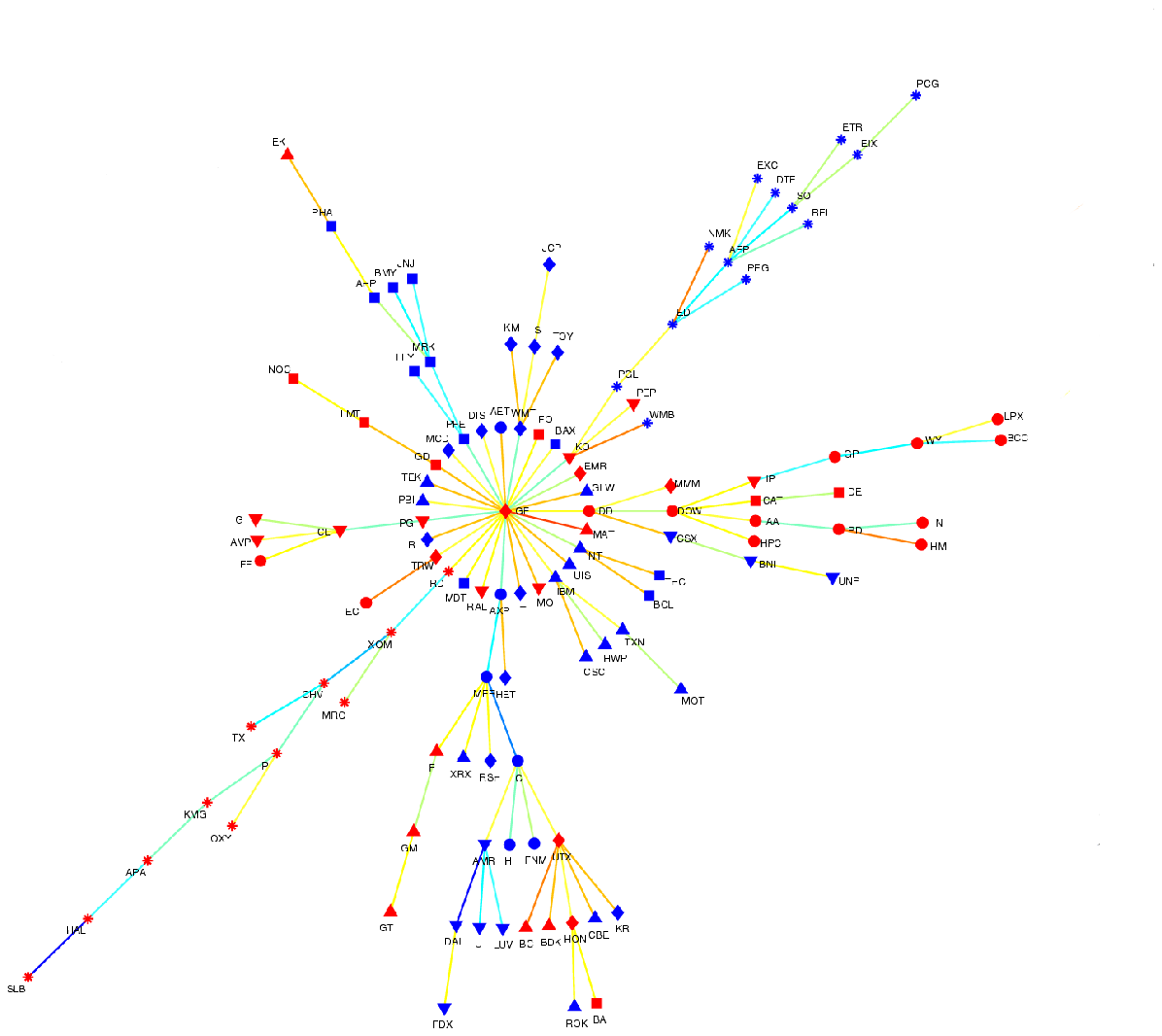}
\includegraphics[width=0.6\textwidth]{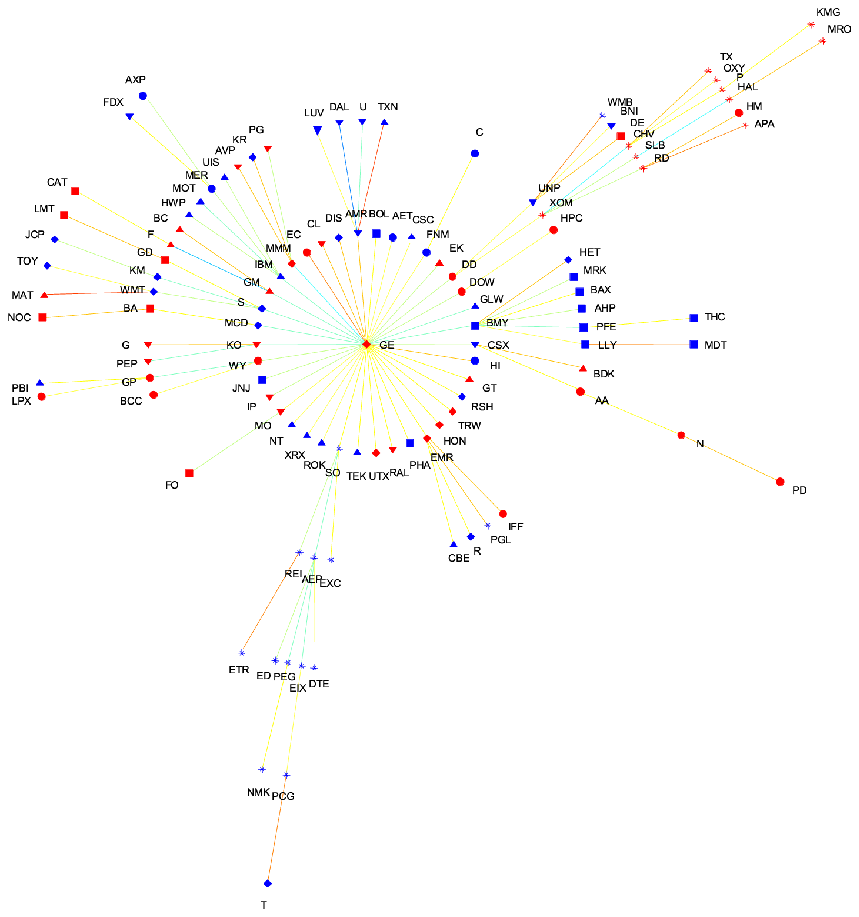}
\end{center}

\caption{Example of a normal (top) and crash (bottom) topology.}
\end{figure}

\end{document}